\documentclass[nofigure]{pasj00}
\usepackage{graphicx}
\usepackage{txfonts}
\Received{$\langle$reception date$\rangle$}
\Accepted{$\langle$acception date$\rangle$}
\Published{$\langle$publication date$\rangle$}
\SetRunningHead{Astronomical Society of Japan}{Usage of \texttt{pasj00.cls}}

\usepackage{times}

\begin{document}

\title{SU Uma stars: Rebrightenings after superoutburst}

\author{Friedrich \textsc{Meyer}, \altaffilmark{1}
        Emmi \textsc{Meyer-Hofmeister}, \altaffilmark{1} }
\altaffiltext{1}{Max-Planck-Institut f\"ur Astrophysik, Karl-
     Schwarzschildstr.~1 \\D-85740 Garching, Germany}

\email{emm@mpa-garching.mpg.de}

\maketitle

\begin{abstract}
SU Uma stars after their long superoutbursts often show single or
multiple rebrightenings. We show how this phenomenon can be understood 
as repeated reflections of transition waves which mediate changes
between the hot and the cool state of the accretion disk and travel
back and forth in the outer disk region, leaving an inner part
permanently hot. This points to a temporarily increased viscosity,
possibly related to the formation of large-scale and longer persisting
magnetic fields by the dynamo operation during the long superoutburst.
The 'mini-rebrightenings' in the early post-outburst light curve of V585 Lyr discovered by Kato and Osaki in {\it{Kepler}} \rm{observations seem to be understandable as a small limit cycle of low luminosity changes originating from a wiggle-feature in the thermal equilibrium curve of the cool optically thick disk.}
\end{abstract}  

\SetRunningHead{F. Meyer and E. Meyer-Hofmeister}
               {Rebrightening in WZ Sge stars}

%

\section{Introduction}

The SU UMa stars, a subgroup of dwarf novae, have
long-lasting and bright superoutbursts with regular dwarf nova
outbursts in between. A further subgroup of these, the WZ Sge stars,
show only long-lasting
and very bright superoutbursts. For both kinds of sources, mainly for
WZ Sge stars, rebrightenings are observed right after the decline from a
superoutburst. During the rebrightening phase repetitive brightness 
variations of one to three magnitudes, lasting several days, are found.
The light curves show an interesting pattern of slightly
steeper rise and slower decay. This brightness variation may be the
mark of the two kinds of transition waves mediating the change
of an accretion disk from cold to hot and from hot to cold state. Such
a pattern of steeper rise and slower decay can also be seen in
light curves of ordinary dwarf nova outbursts (e.g. see Cannizzo et al. 2012).

The so-called 'mini-rebrighenings' in
the \it{Kepler} \rm{ observations of the superoutburst of V585 Lyr,
found and discussed by Kato and Osaki (2013), marginally also show this pattern. These repetitive
semi-periodic brightness variations of 0.4-0.5 mag, with a period of
0.5d, lasting about 4 days, are a new feature
of dwarf nova outburst behavior. V585 Lyr was classified as a
borderline object between SU UMa and WZ Sge-type dwarf novae 
(Kato and Osaki 2013). Such light variations as recorded by the
{\it{Kepler}} \rm{ 
satellite in January-February 2010, have never been
documented before. In  Fig.\ref{f:V585} the light curve
is shown. As Kato and Osaki (2013) pointed out the amplitudes appear to
be too large to be explained by a beat between some period and another,  
and the periods appear to be too short to be explained  by a beat phenomenon, so
the feature should be seen as a real brightness variation with these timescales.

Rebrightenings are observed in several SU UMa stars. Especially the
observation of six consecutive rebrightenings of EG Cnc after its main outburst
in 1996/97 (Patterson et al. 1998) raised the question how these
brightness variations could come about. Hameury (2000), Hameury et al. (2000) and Buat-M\'enard and Hameury (2002) suggested that the rebrightenings could be
explained by enhanced mass transfer due to illumination of the secondary star.
But the investigation of mass transfer lead to the result that the
mass transfer enhancement is neglegible:
(1) the critically examined observations do not show enhanced
mass transfer and (2) the theoretical analysis shows the problems with
an irradiation effect, the shadow of the accretion disk cast on the
$L_1$ point, the high opacity in the extreme ultraviolet band (which
prevents the bulk of radiative flux from reaching the stellar
photosphere) and the strong Coriolis forces (which prevent the heated
material from moving towards $L_1$) (Osaki \& Meyer 2003, 2004). The
analytic result was confirmed in a numerical way by Viallet \& Hameury
(2008).

Osaki et al. (2001) explained the rebrightenings of EG Cnc as
originating from increased viscosity. If the dynamo created magnetic fields in the accretion disk
are responsible for the viscosity and stay on for a while when the
dynamo action has already ended, as magneto-hydrodynamic simulations show
(Balbus and Hawley 1991), then an outburst created increase of the
viscosity also can stay on for some time after the superoutburst and during
that period lead to a sequence of brightness variations until
hot and cold state viscosities have resumed their normal values and the
regular outburst behavior is established.
The mini-rebrightenings are different from the regular
rebrightenings with respect to their rapid repetition, their small
amplitude and the occurrence at almost quiescence luminosity.

In this paper we suggest that the many features in the regular 
rebrightenings can be understood in the context of the general model
of enhanced viscosities at the end of
the superoutburst. Our aim is not the modeling of rebrightenings
with the help of detailed computer simulations, but to show how this
phenomenon might be understood as originating from  
properties of the viscosity related to the disk magnetic
fields, generally accepted as the cause of the disk viscosity.

 \begin{figure}
 \begin{center}
   \includegraphics[width=8.cm]{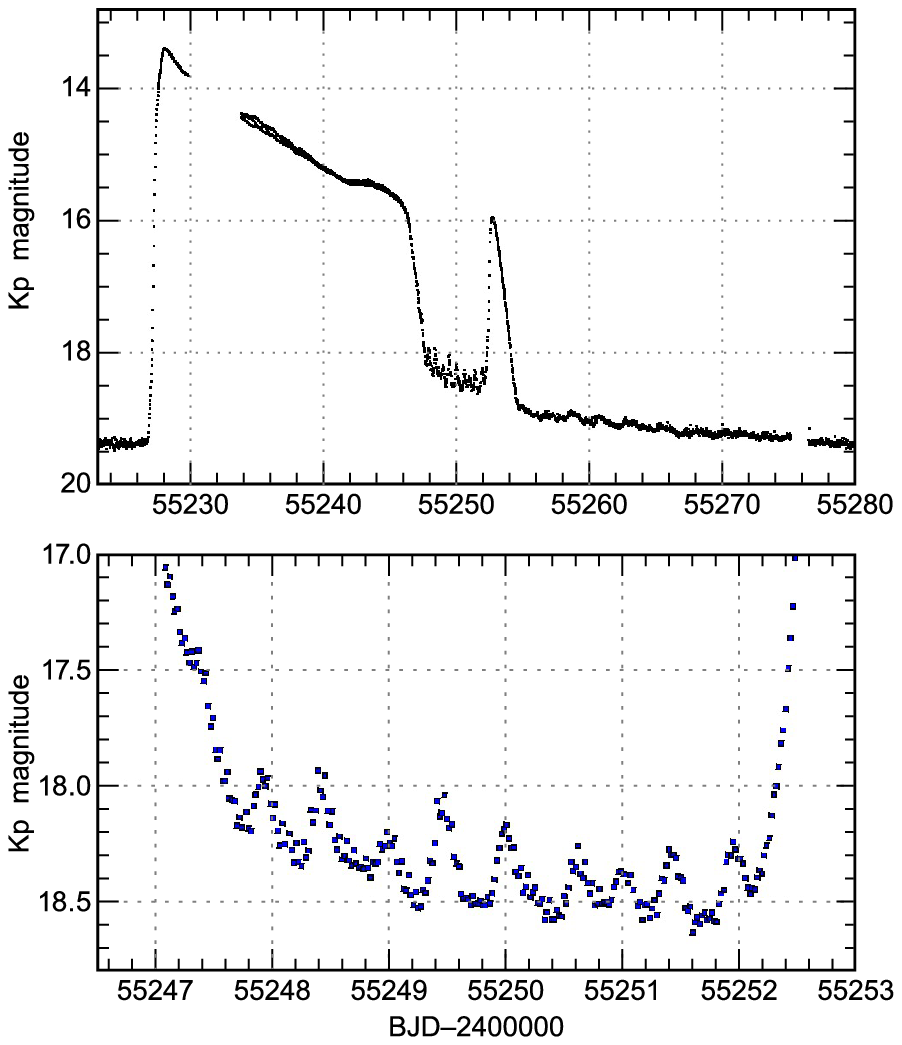}
  \end{center}
       \caption{ The 2010 superoutburst of V585 Lyr in {\it{Kepler}}
         \rm{LC data,
         upper panel: entire outburst, lower panel: post-superoutburst
       state with mini-rebrightenings; from Kato and Osaki (2013).} }
         \label{f:V585}
   \end{figure}

In Section 2 we give a short summary of the physics of dwarf nova 
outbursts and the underlying mechanism of the cyclic changes between a
cold unionized and a hot ionized state of the accretion disk. In
Section 3 we give a short discussion of features of disk magnetic fields and the
consequences for the viscosity they generate. In Section 4 and 5 we report
on the observations of rebrightenings and discuss how the 
rebrightenings can be interpreted as resulting from transition waves
changing the disk
structure back and forth between hot and cool state, propagating
in a limited part of the disk. In Section 6 we discuss a possibility how
the mini-rebrightening phenomenon could be understood. Discussion
and conclusions follow in Section 7 and 8.

\section{Disk structure and dwarf nova outbursts}
The evolution of the surface density distribution depends
on the mass flow and angular momentum transport in the disk and the mass
of the central star. The angular momentum transport is determined by
friction, traditionally parametrized
as the ratio $\alpha$ of viscous stress to gas pressure (Shakura and Sunyaev
1973), its physical cause now seen in the magneto-rotational
turbulence documented by many MHD simulations. The surface density is
determined by vertical structure computations. Fig.\ref{f:fs} shows an
example of the relation between surface density and mass flow rate (or
effective temperature, or emitted flux). The characteristic S-shape of
the equilibrium solution is due to hydrogen ionization in this
temperature range. In this region for the same surface density a hot, high
mass flow rate stable solution and a cool, low mass flow rate stable 
solution exist, with an unstable part in between. The equilibrium
solution curve, here is shown for $r=10^{10}$cm, the shape is about
similar for all the distances from the central star.

We show the equilibria solutions for the two viscosity values 0.05 and 0.2
(for all examples shown here we used the vertical structure results put
together by Ludwig et al. (1994)).
The existence of an unstable branch causes relaxation oscillations,
the dwarf nova outburst cycles.
Already in the beginning of modeling of the observed dwarf nova
outbursts it became clear that for a proper
description of the outburst it is necessary to assume a lower
viscosity value, by about a factor of 4, for the cool state than for the hot
state, e.g. $\alpha_{\rm{cool}}$=0.05, $\alpha_{\rm{hot}}$=0.2 (Smak
1984). The features of the disk instability were modeled 
in many investigations (Meyer and
Meyer-Hofmeister 1981, 1984, Smak 1982, Cannizzo et al. 1982,
Mineshige and Osaki 1983, for a review see Cannizzo
1993, Osaki 1996, Hameury et al. 1998). Comparison of models with
observations including  recent {{\it{Kepler}}} \rm{light curves
(Cannizzo et al. 2012) allows to understand detailed properties of the different
dwarf nova systems. }

During the disk evolution hot and cool state cannot exist 
side by side, instead transition waves move across the
bi-stable region and there mediate the change from the cold to the hot
state and back (Meyer 1984). If a transition wave
reaches the boundary of a bi-stable region it cannot continue and is
reflected as a transition wave of the opposite type. In a standard dwarf nova
outburst cycle (see e.g. Meyer and Meyer-Hofmeister 1984, Ludwig and Meyer
1998) mass accumulation in quiescence leads to a general increase of
surface density at all distances until somewhere in the disk
the critical surface density $\Sigma_{\rm{B}}$ is reached,
and the transition waves spread outward and inward, inducing the
change to the hot ionized state in the whole disk. In the hot disk the mass
flow is high, the surface density decreases until the critical value 
$\Sigma_{\rm{A}}$ is reached at the outer edge of the disk and with
the propagation of the cooling wave the whole disk transits to the cold
state, appearing as quiecence. 

Fig.\ref{f:sasb} shows two examples of disk evolution. For the
first computation standard viscosity values $\alpha_{\rm{hot}}$=0.2, 
$\alpha_{\rm{cool}}$=0.05 were used, for the second 
enhanced viscosity values
$\alpha_{\rm{hot}}$=0.4$,\alpha_{\rm{cool}}$=0.2 (white dwarf mass
1$M_\odot$, outer disk edge at $10^{10.3}$cm in all computations).
Fig.\ref{f:sasb}, top panel, shows surface density distributions for the
standard case, first at the moment when during decline from outburst 
$\Sigma_{\rm{A}}$ is reached at the outer disk edge (upper
curve). The second curve describes the distribution after 
the surface density had at the outer edge decreased to
$\Sigma_{\rm{A}}$, a cooling wave propagated inward and reached the
inner edge. Then the disk is completely cool.  
The bottom panel shows the surface density distribution for the
enhanced viscosity values, also during outburst decline after
$\Sigma_{\rm{A}}$ was reached and the cooling
wave propagated inward. With the enhanced values the critical value
$\Sigma_B$ is closer to $\Sigma_{\rm{A}}$ and the cooling
wave reaches $\Sigma_B$ before the whole disk becomes cool. Since for surface
densities higher than $\Sigma_B$ no cool state exists anymore, the cooling
wave is reflected as an outward propagating heating wave, and the
brightness increases again leading to a rebrightening of the
disk. The further disk evolution is discussed in Section 5. 

These two examples of disk evolution were computed with our code (Meyer and
Meyer-Hofmeister 1984) with separately computed transition wave
velocities (Ludwig et al. 1994). With the localized
front approximation in this code, the decrease of surface density in the
hot region near the transition is not resolved. But at which time
the critical surface density $\Sigma_{\rm{B}}$ is reached is
practically independent of this feature (compare Ludwig and Meyer 1998).}\rm{

 \begin{figure}
   \centering
   \includegraphics[width=7.cm]{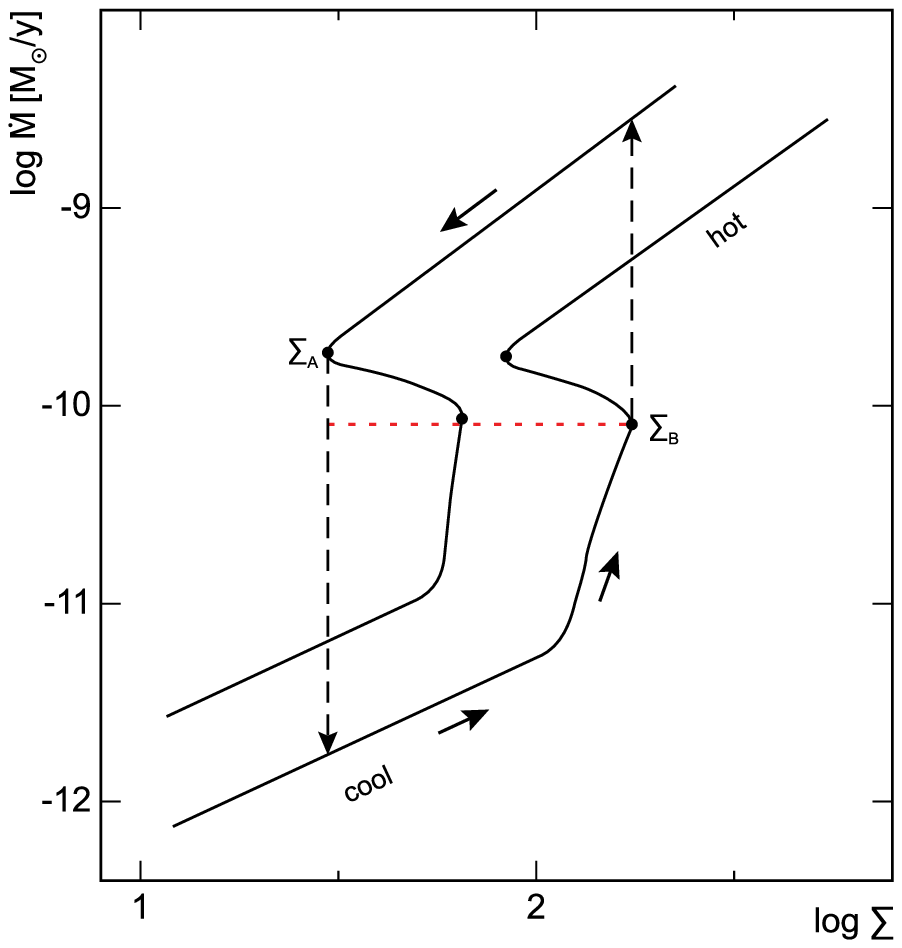}
       \caption{S-shaped thermal equilibrium curves for the two
        viscosity values $\alpha$=0.2 and
         $\alpha$=0.05 (left and right curve). For standard modeling
         of dwarf nova outbursts the combination $\alpha_{\rm{hot}}$=0.2 and 
         $\alpha_{\rm{cool}}$=0.05 is used, see text.
         The curve shows the relation between mass flow
         rate $\dot M$ (corresponding to temperature and vertical
         viscosity integral) and
         the surface density $\Sigma$ at the distance $r=10^{10}$ cm for a
         1 $M_{\odot}$ white dwarf. The
         arrows indicate the cyclic change during an outburst.
         The difference between $\Sigma_{\rm{A}}$ and 
         $\Sigma_{\rm{B}}$ is important for the cyclic behavior.}
         \label{f:fs}
   \end{figure}

}\rm{
Whether reflection of the cooling wave occurs}\rm{ depends on the
difference between the values $\Sigma_{\rm{A}}$ and $\Sigma_{\rm{B}}$
(marked in Fig.\ref{f:fs} by a the dotted red line), and on the
transition wave velocities, which all depend on the viscosity
values taken.

 \begin{figure}
   \centering
     \includegraphics[width=8.cm]{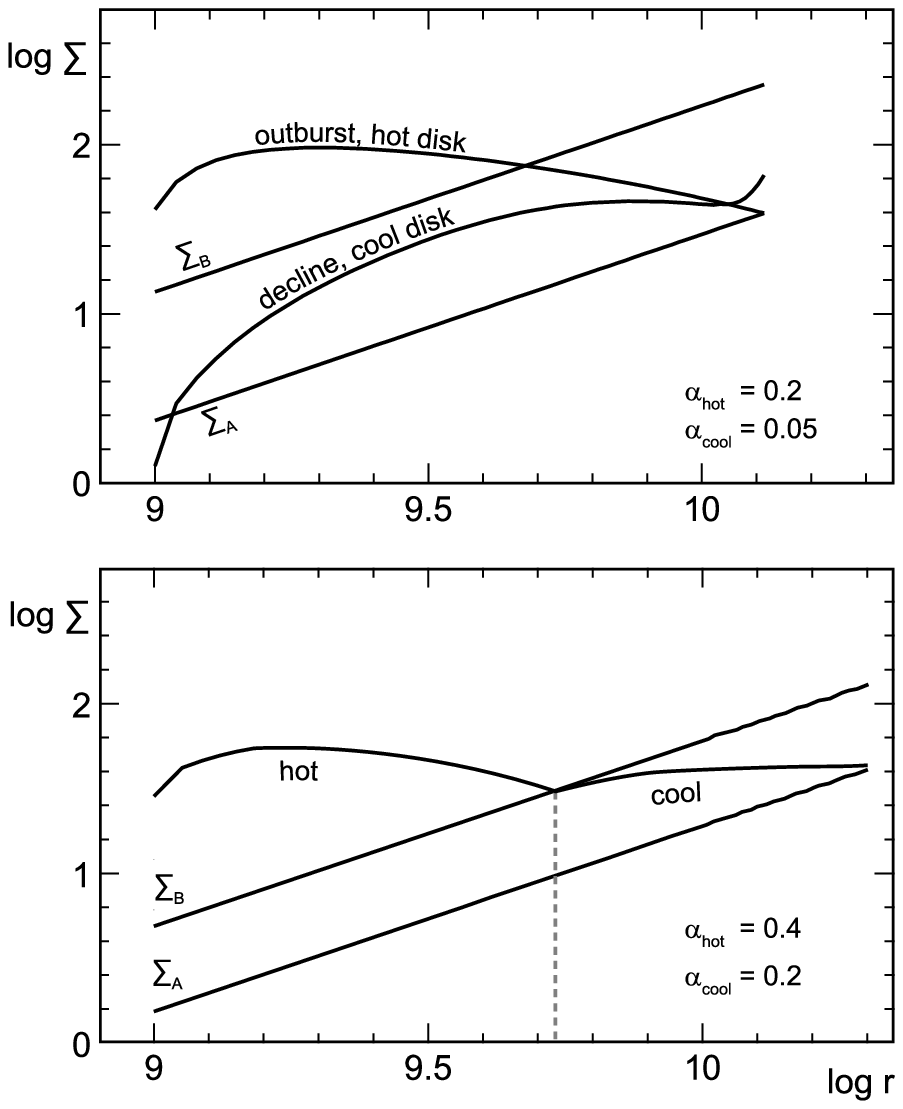}
       \caption{Evolution of surface density $\Sigma$ as function
           of distance $r$, with critical values $\Sigma_{\rm{A}}$ and 
           $\Sigma_{\rm{B}}$.  
                Top panel: Standard dwarf nova
                outburst cycle. Upper curve: in outburst decline when the
                transition to the cool state begins; lower curve: 
                distribution at the end of the outburst decline, when
                the whole disk has become cool; mass overflow rate from the
                companion star $10^{15.7}$ g/s.
                Bottom panel: 
                Distribution during outburst decline for
                the case that the inward propagating cooling wave
                reaches $\Sigma_{\rm{B}}$ before the whole disk
                has become cool, and the cooling wave is reflected
                (note that in the local front approximation used here
                the decrease of surface density in the hot region in
                front of the transition is not resolved, see text). 
                Mass flow rate $10^{15.9}$ g/s.}
           
         \label{f:sasb}
   \end{figure}

For SU UMa stars the complete outburst cycles including the superoutbursts
were explained by a combination of thermal instability with the tidal
instability, the tidal forces caused by the 3:1 resonance between disk and
binary orbit (Osaki 1989).

\section{Viscosity, magnetic fields and dynamos}

In this section we 
are interested in the aspect that the magnetic origin of viscosity
introduces into the accretion physics of dwarf nova outbursts,
in particular for the rebrightening phenomena during the outburst 
cycles of SU Uma stars. 

Even before Balbus and Hawley (1991) discovered the fundamental importance
of the magneto-rotational instability (MRI) for the dynamo generation
of magnetic fields and the viscosity in accretion disks Smak (1984)
pointed out that in order to obtain good DN outburst simulations
the viscosity must differ between hot and cool
state of the accretion disk. 

Now many MHD simulations support the generally accepted concept that
dynamo-created magnetic fields are responsible for the high
viscosity in the outburst. The MRI in the disk is a small-scale
phenomenon that operates on magnetic fields of the scale of the disk
height producing flux bundles of opposite polarity. These flux
bundles connect neighboring regions and exchange angular momentum
between them. Magnetic buoyancy leads to rising and sinking of flux
tubes. The whole process is turbulent. As magnetic flux penetrates
the disk surface one obtains locally separated regions of positive
and negative polarity. The turbulent character constitutes
diffusivity for magnetic fields and angular momentum.

In quiescence in
ordinary dwarf novae electrical conductivity becomes too poor to allow 
dynamo action (Gammie and Menou 1998). But the required
viscosity can now still be provided by the magnetic fields of the
secondary star that penetrate the disk or are entrained with the
accretion, since even at the lower temperature of quiescence the conductivity is
high enough to give a significant coupling between matter and 
magnetic field. In WZ Sge stars the large accumulation of matter
during the extremely long quiescence indicates an
even lower viscosity, e.g. in WZ Sge itself a value
of order 0.001 (Smak 1993, Meyer-Hofmeister et al. 1998). In this
binary during the long lasting evolution the companion star has become a 
degenerate brown dwarf, which neither can operate a stellar dynamo nor hold
magnetic fields from earlier dynamo action, so that this source of
quiescence viscosity disappears (Meyer and Meyer-Hofmeister 1999).

The concept of dynamo-created magnetic fields brings interesting
aspects into the scheme of accretion disk viscosity. The continuous
local dynamo generation and decay of magnetic flux and its diffusion
can lead to large-scale components in the radial distribution. As
shown by MHD simulations of turbulent plasma by Brandenburg (2000)
these large-scale components are built up on their respective
diffusive timescale ('inverse cascade'). With their appearance also
the local saturation field strength increases.
This means an increase
of the local friction and angular momentum transport, in the 
alpha-parametrization an increase of the alpha value. Thus in the course
of a long lasting outburst large-scale magnetic flux distributions 
stochastically evolve (Park 2014). The complete picture is
complex and also includes vertical differences in the formation and
transport of magnetic fields (e.g. Suzuki and Inutsuka 2014). 

When the dynamo action stops and the disk returns to the cool state, 
MRI and turbulence remain since the electrical conductivity is still
sufficiently high to couple matter and magnetic fields, though not
sufficiently high to operate a dynamo. Thus no new flux is generated
and the existing flux distributions will decay on their turbulent
diffusive timescales by mixing of positive and negative polarities. 
For the locally generated fields of size of the order of a few scale
heights the decay time then is very short, about a few times the
Kepler orbital time divided by alpha, about hours. But the components of
the large-scale polarity distribution can only be annihilated on the long large-scale diffusion time, which is the diffusion time for the whole disk, the characteristic time of outburst decline, of the order of weeks.

This character has consequences for the formation of the viscosity in short
and long lasting outbursts. While in ordinary outbursts the matter is
exhausted in one diffusion time of the hot disk, in the long
lasting superoutbursts the flux distributions stay on for much longer time. This
character is mirrored in the magnetically induced viscosity after the
outburst. Also the viscosity for the cool state remaining after a
superoutburst can be significantly enhanced, even more than the hot state
viscosity, because of the additional presence of those dynamo-created fields in quiescence. This has the
effect that the critical values log($\Sigma_A$) and log($\Sigma_B$) become
closer together, in contrast to the situation in ordinary outbursts 
where the dynamo fields have quickly disappeared. This is an important feature
for either standard DN outbursts or the occurrence of reflection of
transition waves. 

For our modeling of rebrightenings we get the picture of increased
alpha-values in long duration superoutbursts while persisting over weeks.

The idea of an enhanced cool state viscosity was already
used in the modeling of the rebrightenings of EG Cnc (Osaki et al. 2001). 
The emerging picture of accretion disk dynamo processes and
magnetic viscosity now opens the possibility to understand the
rebrightenings of the WZ Sge stars in their many different
aspects.

\section{Rebrightening observations}
The rebrightenings or 'echo outbursts' here to be discussed are only 
observed in SU UMa stars,
a subgroup of dwarf nova, which besides the regular outbursts show the 
more luminous and longer duration superoutbursts. The number of regular
outbursts between the superoutbursts decreases with the
orbital period until finally the lowest period WZ Sge
stars only show long-lasting superoutbursts. After
the luminosity decline at the end of a superoutburst sometimes a short 
rebrightening, an increase of luminosity by to 2-3 magnitudes, lasting only a
few days, is observed. Such single or repetitive rebrightenings differ from
regular outbursts in the way that the rebrightening is shorter, the
brightness increase is smaller, but are followed by the usual long
quiescence. Imada et
al. (2006) classified the rebrightenings by their light curve shapes as
the types long-duration, multiple and single. One source 
can display different types after different superoutbursts, as 
e.g. observed for UZ Boo, EZ Lyn and WZ
Sge (Arai et al. 2009, Nakata et al. 2013), which might be interesting 
with respect to the understanding of the physics
involved. Information on rebrightenings is summarized in the surveys by Kato et
al. (2009, 2010, 2012, 2013, 2014), in the recent papers
by Mr\'oz et al. (2013) on OGLE objects, and by Nakata et al. (2013) on MASTER
OT objects. Generally in quite a number of sources indications of
rebrightening are found, but often due to only few observations or intrinsic
faintness the detailed features cannot be seen clearly.

\begin{figure}
\centering
\includegraphics[width=7.cm]{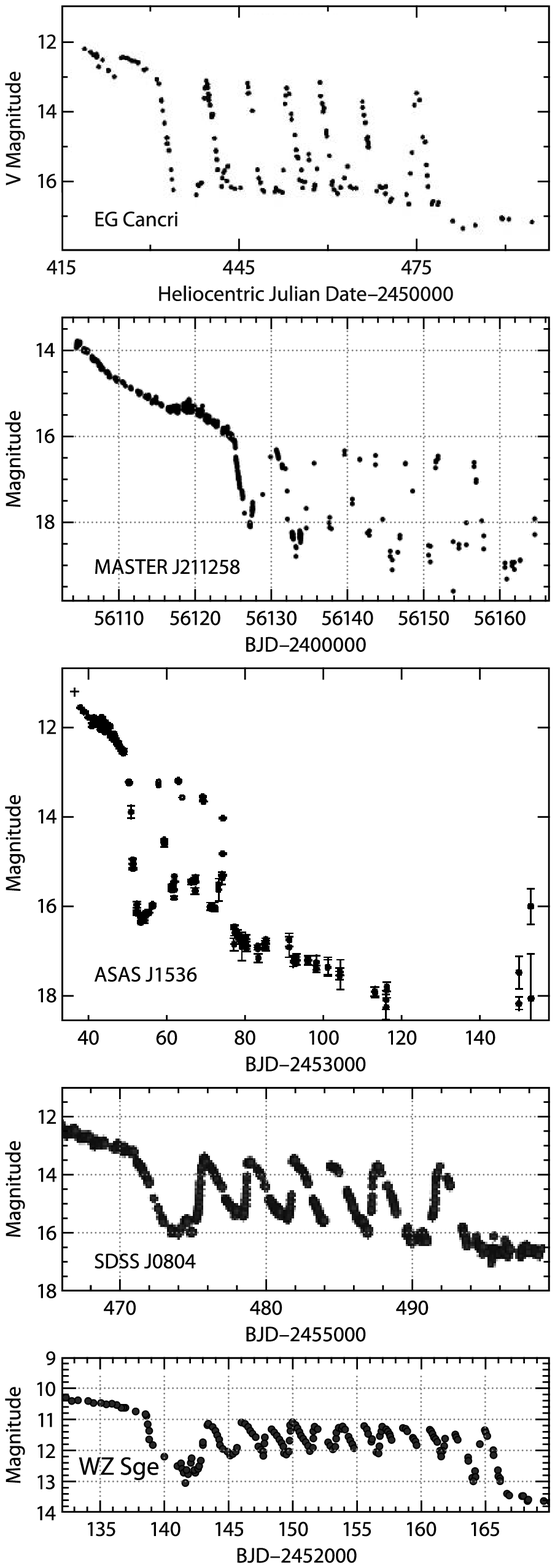}
       \caption{
      Examples of multiple rebrightening: light curves of EG Cnc
      (1996) from Patterson et al. (1998); MASTER J211258 (2012) from Nakata
      et al. 2013; ASAS J1536 (2004) from Kato et al. 2009; SDSS J0804
      (2010) from Kato et al. 2012; WZ Sge (2001) from Patterson
      et al. 2002 and Kato et al. 2009.} 
         \label{f:rebr}
\end{figure}

In Fig.\ref{f:rebr} we show various patterns of multiple
rebrightenings, a series of individual brightness changes of the same
amplitude, following the superoutburst in SU UMa stars.
In EG Cnc the sequence of the rebrightenings resembles that of regular
outbursts with
short quiescence intervals in between, but with an abrupt end after 40
days. The observations of the source  MASTER J211258.65+242145.4
(hereafter MASTER J211258) show almost
the same brightness variation, also seem to reach the quiescence
level (level after the end of the rebrightenings), but no quiescence
intervals. The rebrightenings in QZ Lib (ASAS J1536) do not seem
to come down to the quiescence level. Such a feature shows more clearly in the
better observed EZ Lyn (SDSS J0804). The light curve for WZ Sge an even smaller
rebrightening amplitude can be seen, the luminosity again not coming
down to the later quiescence level. In all these examples the peak
luminosity appeared to reach that at the beginning of the steep
decline before the end of the superoutburst. In stark contrast to
these observations the mini-rebrightenings of V585 Lyr (Kato and Osaki
2013) show luminosity changes of small amplitude at nearly the
quiescence level (Fig.\ref{f:V585}). Such variations had never been
documented before.

In Table 1 we list sources with a relatively
good documentation of multiple rebrightenings: The number of
rebrightenings $N_{\rm{reb}}$ (for some sources only an estimate), 
the derived average duration $t_{\rm{reb}}$  of one individual
rebrightening and the average brightness variation
$\Delta m$ during the rebrightening phase. The amplitude of the individual
brightness change varies somewhat during the rebrightening phase. Our
average values should only be taken as typical values for the
objects. For MASTER J211258 it seemed more probable that only 7
rebrightenings happened (compared to 8 rebrightenings sometimes listed
in the literature). For  OT J012059.6+325545 (hereafter OT
 J0120) where apparently not all minima were
clearly documented (Kato et al. 2012, Fig. 52) we took $\Delta m $=1. 
With respect to the different appearance of the rebrightenings of EZ
Lyn in 2006 and 2010, and similar behavior for other sources, we note
that in the model of magnetic field produced viscosity this feature
could be attributed to the stochastic nature of the dynamo process.
The mini-rebrightenings of V585 Lyr occur on a very short timescale.

\begin{table}
\caption{\bf{ WZ Sge type dwarf novae with multiple rebrightenings.}} 
\label {table:1}
\begin{center} 
\begin{tabular}{lrcccc}
\hline 
Object & Year & $N_{\rm{reb}}$ & $t_{\rm{reb}}$ & $\Delta m $ &
Ref. \\
&&& [d] & [mag] &\\
\hline
DY CMi (OT J0747)& 2008 & 6 & 4.7 & 2.9 & 1\\
EG Cnc & 1996 & 6 & 7.3 & 3.3 & 2\\
VX For & 2009 & 5 & 4.5 & 2.9 & 3\\
QZ Lib (ASAS J1536) & 2004 & 4 & 5.5 & 2.8 & 1\\
EZ Lyn (SDSS J0804) & 2006 & 11 & 2.7 & 1.6 & 4\\
EZ Lyn (SDSS J0804) & 2010 & 6 & 4.9 & 2.5 & 5\\
V585 Lyr & 2010 & 9 & 0.5 & 0.4 & 6\\
WZ Sge & 2001 & 12 & 2.1 & 1.2 & 1\\
OT J0120 & 2010 & 9 & 1.9 & 1.0 & 7\\
MASTER J211258 & 2012 & 7 & 4.7 & 2.6 & 8\\ 
\hline
\end{tabular} 
\end{center}

$N_{\rm{reb}}$ number of rebrightenings\\
$t_{\rm{reb}}$ and $\Delta m$ average values of duration of the
    rebrightenings and brightness variation 
\\
References:(1) Kato et al. (2009), (2) Patterson et
  al. (2002), (3) Kato et al. (2010), (4)  Kato et al. (2009a), (5) Pavlenko et al. (2007), (6) Kato and 
Osaki (2013), (7) Kato et al. (2012), (8) Nakata et al. (2013) 
\end{table} 
\hspace{1.cm}

\section{Rebrightenings: Transitions between hot and
  cold state}

We interpret the appearance of rebrightenings as due to enhanced
viscosity possibly caused by the continuous dynamo action during the
long superoutburst.

For standard viscosity values $\alpha_{\rm{hot}}$=0.2, 
$\alpha_{\rm{cool}}$=0.05 (Fig.\ref{f:sasb}, upper panel) the critical
lines $\Sigma_A$ and $\Sigma_B$ are far apart, the inward moving
cooling wave never reaches $\Sigma_B$, and the whole disk becomes cool
and enters into quiescence.

For increased viscosity values $\alpha_{hot}$=0.4, $\alpha_{cool}$ =0.2
(Fig.\ref{f:sasb}, lower panel) however $\Sigma_A$ and
$\Sigma_B$ are closer together, the cooling front at some inner
point reaches $\Sigma_B$ and is reflected as an
outward propagating heating wave, which results in the rebrightening
of the disk. When this heating wave has reached the outer edge of 
the disk, where the surface density distribution peters out to values
below  $\Sigma_A$ (here taken as the 3:1 resonance radius) it is reflected
as a cooling wave. At this moment (almost) the whole disk is in hot
state and the brightness is at its maximum. Where exactly the
reflection point lies depends on how the angular momentum loss by the
3:1 resonance is distributed and would require detailed modeling to be
determined. The whole process then can repeat.
 
But we note that to model the observed rebrightenings with the
approximately same 
brightness maximum during the repetitions a part of the disk has to be in a 
quasi-stationary state, modulated by the repeated rebrightening
and requiring a continuous mass inflow from the outermost regions. This
seems plausible due to the storage of matter beyond the 3:1 resonance
radius (Osaki et al. 2001). Hellier (2001) has pointed out the
large extension and the eccentricity of the accretion disk in EG Cnc
and that the enhanced angular momentum transport from the 3:1
resonance ensures an enhanced mass flow into the inner disk for
the entire period of the superoutburst and consecutive
rebrightenings, documented by the persistent superhumps. The
eccentricity important for the inflow of matter would not lead to
dynamo action since the temperature increase involved appears too low.
(A simple estimate of
the effective temperature resulting from the standard viscous
transport of mass and the additional heating by the tidal transfer of
angular momentum yields $T_{\rm{eff}} \approx 5400\rm{K}({\dot
M_{16}/r_{10.3})^{1/4}}$, with ${\dot M_{16}}$ mass transfer rate in units of
$10^{16}$g/s, $r_{10.3}$ resonance radius in units of
$10^{10.3}$cm and an assumed concentration of the effective dissipation
region to 1/10 of the radius.)

Our model} \rm{interprets the repeated rebrightenings as the multiple
reflections of transition waves in the outer disk area, while an
inner part of the disk remains in the hot state. Dependent on at which
disk radius $r$ the reflection of the cooling wave occurs the
extension of the inner hot region is smaller or larger, resulting
in a lower or higher luminosity at minimum.

The light curves of superoutbursts show at first a gradual decrease
when the whole disk is still hot, then a steep decline when the
surface density has decreased to $\Sigma_A(r)$ and the cooling front
propagates inward decreasing the extent of the still hot region. The luminosity
when the steep decline begins is the same as the maximum
luminosity during rebrightenings. In our computation the luminosity
decrease, from $10^{33.13}$ erg/s (completely hot disk) to $10^{32.4}$
erg/s (a still hot inner part at $r \leq 10^{9.73}$cm) (the minimum during
rebrightening), that is the transition of the outer part to the cool
state, occurs within 2.1 days. The heating wave which leads to the
luminosity increase during the rebrightening propagates faster, the
change occurs within less than 1 day. Such brightness
changes of almost 2 magnitudes with a repetition time of
2-3 days would be typical for observed rebrightenings.}\rm{

The enhanced viscosities used in our example for rebrightenings
(present after the long superoutburst) seem in agreement with the 
picture of the magnetic origin of the viscosity discussed in Sect.3.

With this in mind we distinguish the following types of
rebrightening:
\begin {itemize}
\item type a: The
cooling wave can propagate (almost) to the inner edge, therefore the luminosity
is almost as low as in quiescence. This is similar to ordinary dwarf nova
outbursts, except the shorter outburst and the existence of a very
short (or no) quiescence.
\item type b: If the
 critical value $\Sigma_{\rm{B}}$ is reached early
by the cooling wave, a larger inner disk stays hot, consequently the
amplitude of brightness variation is smaller. The quiescence
luminosity, i.e. the luminosity of a completely cool disk, is not reached.
\end{itemize}

These different types are visible in the examples in
Fig.\ref{f:rebr}. The rebrightenings in EG Cnc (1996/1997) show very
short quiescence intervals, type a. In MASTER J211258 (2012) only
variations up and down are visible, probably without quiescence, so
a marginal type a. In the light curve of ASAS J1536 (2004) the
minimum luminosity of the rebrightenings seems to be higher than the
quiescence level. In the following example, SDSS J0804, the brightness 
minimum is above the quiescence level, at least during the early 
rebrightenings. Clearly visible is this effect for WZ Sge , (and is
also the case for the light curve of OT J0120 (Kato et al. 2012), not
shown here), all examples type b.

Another pattern appears in V585 Lyr (Fig.\ref{f:V585}), which stands
out in a much more rapid and low amplitude brightness variation in a
nearly quiescent disk. We come back to this case in the following section.

\begin{figure}
   \begin {center}
     \includegraphics[width=7cm]{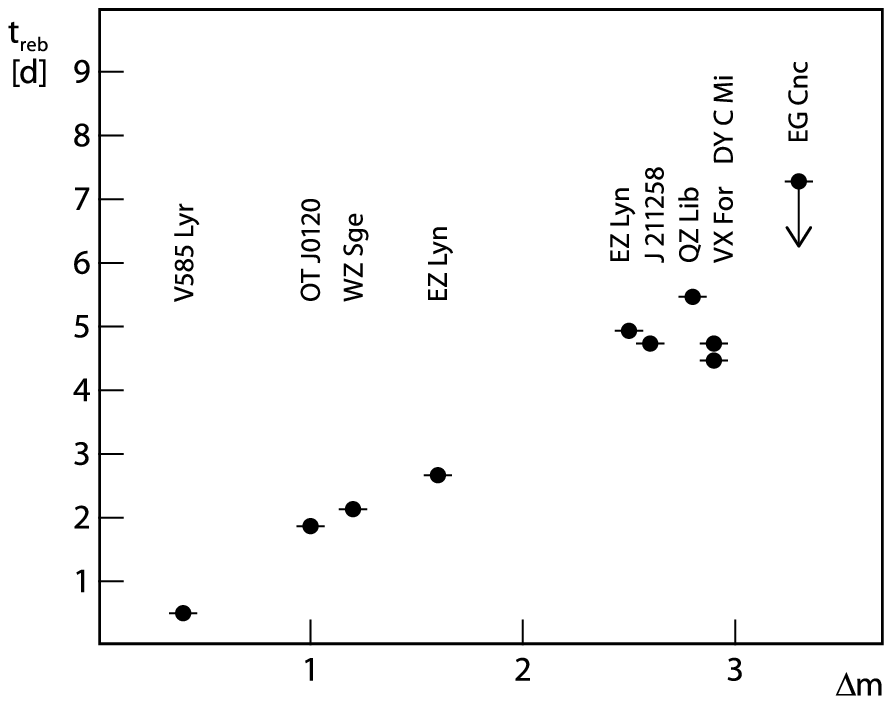}
   \end{center} 
       \caption{Relation between duration of one repetitive
       rebrightening (increase + decrease) $t_{\rm{reb}}$ and
       brightness variation $\Delta m$ for the sources in Table 1}
           \label{f:dur}
   \end{figure}

In these examples of rebrightening, if sufficiently resolved, one can recognize 
the more rapid brightness increase and the slower brightness decrease, 
a sign of the faster moving heating waves and the more slowly moving 
cooling waves. The same pattern is 
present in the rise and decline of regular dwarf nova outbursts
(e.g. in recent \it{Kepler} \rm{light curves, Cannizzo et
  al. 2012). \rm{  }

Fig.\ref{f:dur} shows a relation between $t_{\rm{reb}}$ and $\Delta
m$ (listed in Table 1). In our interpretation
$t_{\rm{reb}}$ is the time needed for the transition waves to propagate through
that region of the disk which changes back and forth between the hot and cool
state, a  larger brightness variation and a longer time for a larger 
region. (For EG Cnc, which has short
quiescent intervals in between the rebrightenings, one should subtract
these quiescent times, when determining $t_{\rm{reb}}$ as indicated by the
arrow.) The proportionality of $t_{\rm{reb}}$ and $\Delta
m$ suggests that the transition wave velocities in all these
rebrightenings are about the same, as theory predicts, supporting this
interpretation.

\section{The mini-rebrightenings}

Kato and Osaki (2013) found and discussed a new feature in the
{\it{Kepler}}\rm{
observations of the superoutburst of V585 Lyr, repetitive
semi-periodic brightness variations of 0.4-0.5 mag, with a period of
0.5d, lasting for about 4 days, as shown in Fig.\ref{f:V585}. As the authors
pointed out the origin of these mini-rebrightenings is not clear.
How can such short-time brightness variations come about? 
Besides their rapid oscillations these mini-rebrightenings are also
peculiar in their low brightness and small amplitude variation.

Already during early dwarf nova outburst modeling Mineshige and Osaki
(1985) for constant viscosity values $\alpha_{hot}$=$\alpha_{cold}$, $\Sigma_A$ and
$\Sigma_B$ close together, had found small amplitude continuous
luminosity variations (one fluctuation lasting 4 days), where the
inner part of the disk remains in hot state, the outer part in cold
state, and the transition fronts travel back and forth in an
in-between unstable region, never turning the disk completely hot or
cold. For the thermal instability of accretion disks in galactic
nuclei Mineshige and Shields
(1990) similarly found fast small variations in the V-magnitude,
called ``purr'' outbursts, using a
viscosity description with critical surface density
values close together, log($\Sigma_B$) - log($\Sigma_A$) = 0.25 (see their
Fig. 2). They described these variations as 
 'caused by heating and cooling fronts which rapidly traverse back and forth
 across the unstable zone'. In both these investigations an inner part
of the disk remains always hot, which results in a luminosity minimum higher
than the quiescence value. 

If this picture should also be applicable to mini-rebrightenings
the very short duration of their individual brightening together with the
velocity of transition waves requires a very small width of the unstable
region. As the mass flow rate in the cold and in the hot state at the two
boundaries would be approximately the same this translates into a very 
small difference between the two critical values $\Sigma_A$ and
$\Sigma_B$. In our case this would require $\alpha_{cold}$ even
significantly larger than $\alpha_{hot}$, an implausible choice of
viscosity. If such a change between a hot and a cold state is
excluded the question arises whether the mini-rebrightening phenomenon
could be caused by a small limit cycle in the cold state alone. This
seems possible.

\vspace {0.5cm}

\subsection{'Wiggles' in the thermal equilibrium curve}

We here consider whether the mini-rebrightenings can be due a negative slope
in a small part in the cold optically thick region of the thermal
equilibrium curve, causing a small limit cycle. The appearance of a
'wiggle' in the lower part of the S-shaped equilibrium curve as shown in
Fig.\ref{f:wiggle} can lead to small amplitude brightness variations
near the quiescent level. This is a similar limit-cycle behavior as
during the large cycle
between hot and cold state, responsible for a dwarf nova outburst, but
at much lower temperatures, with transition waves mediating the
changes between the two states. The velocity of transition waves is of
the order of the sound velocity multiplied by an $\alpha$ value,
leading to a small difference between heating and cooling wave 
velocity, and to a marginally faster rise and a slower decline of brightness.
Indeed in the mini-rebrightenings of V585 Lyr such a pattern might
be  marginally indicated. It is perhaps interesting that the data for
mini-rebrightening duration and brightness variation of V585 Lyr lie
on the correlation track for regular rebrightenings shown in Fig.\ref{f:dur}.
But this needs further evaluation.

\begin{figure}
   \centering
   \includegraphics[width=5cm]{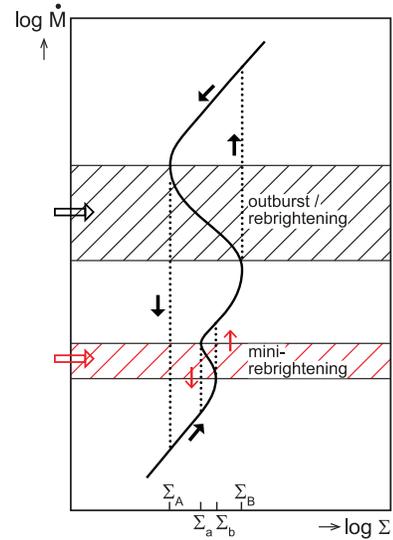}
    \caption{ Schematic drawing of an equilibrium curve
      (Meyer-Hofmeister 1987), with a 'wiggle',
      as present in some computed curves (references in
      text). The small unstable part in the curve, $\dot M$ region
      marked in red,  causes a short-duration, small-amplitude limit
      cycle behavior} 
         \label{f:wiggle}
  \end{figure}

To find out whether wiggles can cause the mini-rebrightenings it 
has to be clarified, (1) under
which conditions wiggles (with an unstable negative slope) are present
in the equilibrium curve, and (2) how
the initiated small limit cycles show up in the light curve.

Wiggles are indicated in a number of vertical structure computations.
Cannizzo and Wheeler (1984) had found
wiggles already in their early work (see their Fig.11). They discussed
in detail how a non-monotonic relation between effective temperature
and surface density can result from an opacity
decrease at T$\sim$ 6000-7000K and the onset of convection. The thermal
equilibrium curve of Mineshige and Osaki (1985, their Fig.2)
clearly has a wiggle
around the effective temperature log$T\approx$ 3.6. The effect of
different low temperature opacities on the shape of the equilibrium
curve was also discussed by Meyer-Hofmeister (1987, Fig.1). The
treatment of convection influences the vertical structure. Besides the
choice of the standard mixing-length parameter especially the way of
limiting the mixing-length itself near the midplane where the pressure
scale height formally becomes infinite can lead to the appearance of an
unstable part in the low temperature optically thick equilibrium curve
Ludwig 1992, Ludwig et al. 1994).

In the investigation of the instability in galactic nuclei Mineshige and
Shields (1990) found equilibrium curves
with slight wiggles (see their Figs.1 and 2). Janiuk et
al. (2004) mentioned the appearance of wiggles in equilibrium curves
for stellar and galactic disks (their Figs.1 and 2).
In the study of the disk instability models for AGN 
Hameury et al. (2009) also found equilibrium curves with wiggles
(their Fig.1). They attributed their appearance to the partial dissociation 
of molecular hydrogen which influences the adiabatic gradient in convection.

Thus the shape of the equilibrium curve, especially a presence of wiggles,
depends on the assumptions and approximations used to
calculate the equilibrium curve, on the opacities used, on
the parametrization of $\alpha$, on the
treatment of convection and assumptions on the mixing length
description. High resolution observations like the }{\it{Kepler}} light curves
of V344 Lyrae might allow to learn about the viscosity values from
modeling the outburst behavior (Cannizzo et al. 2010). 

\subsection{The luminosity variation}

An essential feature of the cold disk during all changes initiated by the
unstable part is the long diffusion time of the cold disk. The
diffusion time $t_{\rm{d}}=(1/\alpha \Omega)(r/H)^2$ with Kepler
frequency $\Omega=\sqrt{GM/r^3}$ and scale height
$H=\sqrt{2\frac{\Re}{\mu}Tr^3 /GM}$, $M$ white dwarf mass, for
T=$10^{3.8}$K, $r$=$10^{9.5}$cm and $M=1 M_\odot$ is
$\approx$30 days. This means that during
the mini-rebrightening phase of 5 days the surface density cannot change
significantly. It also means that the rapid fluctuations in the unstable
band have little effect on the surrounding stable areas and
the luminosity variation is only due to the brightness variation in
the unstable region. The variation there can be roughly estimated
from its range in $r$ and the difference of brightness on the two stable
branches of the wiggles.

When the calculated equilibrium curves at some distance $r$ show a
wiggle with an unstable part it lies in a small range of $\Sigma$
values somewhat below the critical value $\Sigma_{\rm B}$, about the
same for different values of 
$r$, since both features occur at their fairly constant characteristic
temperatures (compare the curves calculated by Mineshige and Shields for AGN,
1990). In a $log\Sigma$-$log r$ diagram then the surface density
values for which wiggles appear lie in a band, roughly parallel to
$\Sigma_{\rm B}(r)$.

Model computations give a shape of the surface density distribution
left over by the inward propagating cooling wave in outburst decline,
a 'common' shape present in outburst computations for different
parameters (an example in our Fig.\ref{f:sasb}, top panel). In order
that a mini-rebrightening phenomenon appears this left over surface
density must reach the wiggle band, but not overshoot too much,
otherwise it would also reach $\Sigma_{\rm B}$ and we would have the
case of standard rebrightening. If the wiggle band and the
$\Sigma_{\rm B}(r)$ line lie close together this means that the left over
density distribution should pass nearly tangentially through the wiggle
band and therefore in a somewhat extended region of $r$. Thus one expects that
the brightness variations occur in a range roughly of order $r$.

The expected surface brightness variation follows from the difference between
$T_{\rm{eff}}^4$ on the two stable branches of the wiggles which we
take here as about a factor of 1.6 (see the calculated curves
referred to above). We note there is almost no change in mass flow
rate, only the temperature changes between higher and lower values.

The region of extension $r$ around $r=10^{9.5}$cm (the 'tangential'
point) with an effective temperature of $10^{3.7}$K radiates a
bolometric luminosity
$L_{\rm{bol}}$=$2\times 2 \pi r^2 \sigma T_{\rm{eff}}^4$=$10^{30.7}$ erg/s
(both sides of the disk), $\sigma$ Stefan-Boltzman constant. A bolometric
correction of 0.4 magnitude leads to a visual luminosity
of about $L_{\rm{vis}}= 10^{30.5}$erg/s. By this amount the light
would fluctuate.

In order to obtain the corresponding variation in magnitude we need
the luminosity of the whole accretion disk
in the beginning quiescence. As the distance to V585 Lyr is not known
we cannot use the observed apparent magnitude to get the absolute
visual luminosity from the observations. Thus for comparison
we give here only some values from model calculations for post outburst
luminosity in some cataclysmic variables. In the
model computations of Meyer and Meyer-Hofmeister (1984) the
distribution of the bolometric luminosity in the cold disk after
outburst decline is shown, with a luminosity of $10^{30.8}$ erg/s for a disk size
of $10^{10.3}$cm, and visual luminosities of the total disk in the
range of $10^{30.3}$ erg/s for OY Car, and $10^{31}$ erg/s for U Gem.
Typical values found by Mineshige and Osaki (1985) for
a relatively large disk are $L_{\rm{bol}}$=$10^{31.3}$ to
$10^{31.5}$ erg/s). A brightness variation of half a magnitude as observed
for V585 Lyr would correspond to a visual luminosity of $10^{30.8}$erg/s and lie
within this very wide range.

\subsection{Repetition rate of the mini-rebrightenings}
The mini-rebrightenings in Fig.\ref{f:V585} repeat in about
$10^{4.7}$s. During this time the transition waves should have
traveled back and forth through the unstable region, extending about
$10^{9.5}$cm. Taking their velocity as $\alpha$ times the sound
velocity, where $\alpha$ is a  mean value of $\alpha$ parameters for
vertical thermal relaxation and radial diffusion in the transition
fronts, a sound speed of $10^{5.8}$cm/s (T=$10^{3.8}$K, neutral
hydrogen), $\alpha$ would be 0.2, a perhaps not unreasonable 
cold $\alpha$ value after the long superoutburst.

These estimates for both, the amplitude and duration of the
fluctuations, indicate that the appearance of wiggles could be a
promising candidate for the understanding of mini-rebrightenings.

\subsection{Mini-outbursts in V1504 Cygni}

We mention here that the occurrence of a limit cycle smaller than
the large cycle in standard outburst modeling, though not as small
as discussed here for the mini-rebrightenings, was already suggested
by Osaki and Kato (2014) for the explanation of 'mini-outbursts',
oubursts of small amplitude, found in the recent {\it{Kepler}}
light curve of the SU Uma star V1504 Cygni appearing in the second
half of a supercycle. In this model the structure of the accretion
disk changes between the cold state and an intermediate warm
state. Such an intermediate stable state can be seen in the thermal
equilibrium curve computed by Mineshige and Osaki (1985, Fig.2). Osaki
and Kato (2014) attributed the appearance of this feature to a
possible additional heating by increased tidal interaction.

\section {Discussion}

\subsection{The probability of occurrence of rebrightenings}

One might ask, how often the superoutburst caused magnetic field
structure leads to a viscosity enhancement that brings about
rebrightenings. The observations tell us, that it is not a rare
phenomenon: The first survey of superhumps of SU UMa stars by Kato et
al. (2009) includes several hundred observations of SU UMa stars, of
which 34 WZ Sge stars are listed. In 20 of those sources 
rebrightenings were found, about half of them multiple. This indicates
that the viscosity after the superoutburst is frequently enhanced.

To obtain a full MHD picture of these derived features will require a
global accretion disk simulation with high spatial and time
resolution, not yet in reach.

\subsection{A relation between superoutburst duration and viscosity
  enhancement ?}

If during a long lasting superoutburst large-scale magnetic fields
are generated and persist for longer time after the end of the
superoutburst the ensuing viscosity enhancement stays on longer,
leading to a larger number of rebrightenings. If the viscosity
increase also becomes larger by the superposition of large-scale
magnetic fields the reflection of the cooling wave occurs earlier and
the region always staying hot becomes larger, i.e. the minimum
brightness is larger.

Such a behavior would be expected predominantly in WZ Sge
stars with long lasting outbursts. This is indeed true for the
examples shown here: the longest rebrightening phase and a higher
minimum brightness are found for WZ Sge itself.

\section {Conclusions}

Accretion disks with mass flow rates as usually present in cataclysmic
variables can be in a hot or a cool state, corresponding to the high
luminosity in a dwarf nova outburst and the low luminosity in
quiescence. The change between hot and cool state is mediated by
transition waves which propagate across the disk. In standard dwarf
nova outbursts and in rebrightenings the steeper rise and the
more slow decline of luminosity is a clear indication of faster moving
heating waves and more slowly moving cooling waves.

We suggest that the various forms of rebrightening in SU UMa stars,
occurring only after superoutburst, can be understood as the
consequences of enhanced viscosities left over from the
superoutburst. They affect the critical surface
densities $\Sigma_{\rm{A}}(r)$ and $\Sigma_{\rm{B}}(r)$ and thereby
the location and the range of the region where 
transition fronts move. This determines maximum luminosity, amplitude
and duration of the individual rebrightening. 
Our modeling is based on simple computations of disk structure
evolution, but allows to understand the consequences of the
enhancement of viscosities following a superoutburst. Such an
enhancement would be related to the long-time dynamo action during the
superoutburst and could explain why rebrightenings only happen after
superoutburst. The intrinsically stochastic nature of the produced
magnetic field structure could also explain why one and
the same system after different superoutbursts can show a different
course of its rebrightenings. The magnetic origin of the disk
viscosity is in general agreement with the now widely accepted 
magneto-rotational instability modeling.

We consider whether the mini-rebrightenings could also be
understood due to  a limit cycle, but this time caused by small
S-shaped 'wiggles' in the cool optically thick part of the thermal
equilibrium curve. Without more detailed calculations of the outburst
decline our estimates seem to indicate that the short period and the small
amplitude of the mini-rebrightenings could be due to
vthis instability. This would be further insight into a special feature of
the thermal viscous accretion disk instability and will deserve
more detailed investigation.

\vspace{0.2cm}

We thank Hans Ritter for many fruitful discussions and the referee for
interesting, helpful comments.

{} 
}

\end{document}